\documentclass[fleqn]{llncs}

\usepackage{color}

\usepackage{todonotes}
\usepackage{hyperref}
\usepackage{color}

\renewcommand{\tt}{\ttfamily}
\newcommand{\codefont}{\small\tt}
\newcommand{\code}[1]{\mbox{\codefont{#1}}}
\newcommand{\ccode}[1]{``\code{#1}''}
\newcommand{\us}{\raise-.8ex\hbox{-}}

\newcommand{\equprogram}[1]{%
\def\separator{0.7ex}%
\frenchspacing%
\refstepcounter{equation}%
\par\vspace\separator\hspace{-0.5em}%
$\vcenter{\codefont\noindent{#1}}$%
\raisebox{-0.0ex}{\kern-1.20em\llap{\rm (\theequation)}}
\par\vspace\separator\noindent\kern-.0em%
}

\usepackage{listings}
\lstnewenvironment{curry}[1][]
  {\lstset{literate={->}{{$\rightarrow{}\!\!\!$}}3,#1}}{}
\newcommand{\listline}{\vrule width0pt depth1.5ex}

\begin{document}
\pagestyle{plain}
\sloppy

\title{Memoized Pull-Tabbing for\\ Functional Logic Programming}

\author{
Michael Hanus
\kern1em
Finn Teegen
}
\institute{
Institut f\"ur Informatik, CAU Kiel, 24098 Kiel, Germany\\
\email{\{mh,fte\}@informatik.uni-kiel.de}
}

\maketitle

\begin{abstract}
Pull-tabbing is an evaluation technique for functional logic programs
which computes all non-deterministic results in a single graph structure.
Pull-tab steps are local graph transformations to move
non-deterministic choices towards the root of an expression.
Pull-tabbing is independent of a search strategy so that
different strategies (depth-first, breadth-first, parallel) can be used
to extract the results of a computation.
It has been used to compile functional logic languages
into imperative or purely functional target languages.
Pull-tab steps might duplicate choices in case of shared subexpressions.
This could result in a dramatic increase of execution time
compared to a backtracking implementation.
In this paper we propose a refinement which avoids this efficiency
problem while keeping all the good properties of pull-tabbing.
We evaluate a first implementation of this
improved technique in the Julia programming language.
\end{abstract}

\section{Introduction}
\label{sec:Introduction}

Functional logic languages \cite{AntoyHanus10CACM}
combine the main features of functional and logic languages
in a single programming model.
In particular, demand-driven evaluation of expressions
is amalgamated with non-deterministic search for values.
This is the basis of optimal evaluation strategies
\cite{AntoyEchahedHanus00JACM} and yields
a tight integration between specifications and code
\cite{AntoyHanus12PADL}.
However, it also demands for advanced implementation
techniques---an active research area in declarative programming.
This paper proposes a new implementation model
which combines advantages of existing models in a novel way.

The main challenge in the implementation of functional logic languages
is the handling of non-determinism.
For instance, consider the following operations
(in example programs we use Curry syntax \cite{Hanus16Curry}
which is close to Haskell):
\begin{curry}
flip 0 = 1              coin = 0
flip 1 = 0              coin = 1
\end{curry}
\code{flip} is a conventional function whereas \code{coin}
is a \emph{non-deterministic operation} \cite{GonzalezEtAl99},
an important concept of contemporary functional logic languages.
A non-deterministic operation might yield more than one result
on the same input, e.g.,
\code{coin} has values \code{0} and \code{1}
(see \cite{AntoyHanus10CACM,GonzalezEtAl99} for discussions of this concept).
Due to the importance of non-deterministic operations,
Curry defines an archetypal \emph{choice} operation \ccode{?} by
\begin{curry}
x ? _ = x
_ ? y = y
\end{curry}
so that one can define \code{coin} also by \ccode{coin = 0$\;$?$\;$1}.
In functional logic languages,
non-deterministic operations can be used as any other
operation, in particular, as arguments to other (deterministic)
operations, e.g., as in \ccode{flip$\;$coin}.
It is important to keep in mind that any evaluation of an expression
might lead to a non-deterministic choice.
We review some existing approaches
to deal with such choices during program execution.

\emph{Backtracking} implements a choice by selecting one alternative
to proceed the computation. If a computation comes to an end
(failure or success), the state before the choice is restored and
the next alternative is taken.
Backtracking is the traditional approach of Prolog systems
so that it is used in implementations that compile
functional logic languages into Prolog, like
PAKCS \cite{AntoyHanus00FROCOS,Hanus18PAKCS} or
TOY \cite{Lopez-FraguasSanchez-Hernandez99}.
The major disadvantage of backtracking is its operational
incompleteness: if the first alternative does not terminate,
no result will be computed.

\emph{Copying} or \emph{cloning} avoids this disadvantage
by copying the context of a choice and proceed with both
alternatives in parallel or by interleaving steps.
Due to the cost of copying when a choice occurs deeply in an expression,
it has been used only in experimental implementations,
e.g., \cite{AntoyHanusLiuTolmach05}.

\emph{Pull-tabbing} is another approach to avoid the incompleteness
of backtracking by keeping all alternatives in one computation structure,
typically, a graph.
It was first sketched in \cite{AlqaddoumiAntoyFischerReck10}
and formally explored in \cite{Antoy11ICLP}.
In contrast to copying, a pull-tab step is a local transformation
which moves a choice occurring in an argument of an operation
outside this operation. For instance,
\begin{curry}
flip (0 ? 1) $~\to~$ (flip 0) ? (flip 1)  
\end{curry}
is a pull-tab step.
Pull-tabbing is used in implementations targeting complete search
strategies, e.g., KiCS \cite{BrasselHuch07},
KiCS2 \cite{BrasselHanusPeemoellerReck11}, or
Sprite \cite{AntoyJost16}.
Although pull-tab steps have local effects only,
iterated pull-tab steps move choices to the root of an expression.
If expressions with choices are shared
(e.g., by \code{let} expressions or multiple occurrences of argument variables
in rule bodies),
pull-tab steps produce multiple copies of the same choice.
This could lead to unsoundness, which can be fixed by
attaching identifiers to choices \cite{Antoy11ICLP},
and to duplication of computations.
The latter is a serious problem of pull-tabbing implementations
\cite{Hanus12ICLP}.
In this paper, we propose a working solution to this
problem by adding a kind of memoization to pull-tabbing.
With this extension, pull-tabbing becomes faster than backtracking
and at the same time flexible and operationally complete search strategies
can be used.

This paper is structured as follows.
After reviewing some details of functional logic programming
and the pull-tab strategy along with its performance issues
in the following two sections, we present our solution to these
problems in Sect.~\ref{sec:MPT}.
A prototypical implementation of our improved strategy
is sketched in Sect.~\ref{sec:implementation} and
evaluated by some benchmarks in Sect.~\ref{sec:benchmarks}.
Related work is discussed in Sect.~\ref{sec:related}
before we conclude.

\section{Functional Logic Programming with Curry}
\label{sec:Curry}

The declarative multi-paradigm language Curry \cite{Hanus16Curry},
considered in this paper for concrete examples,
combines features 
from functional programming (demand-driven evaluation, parametric
polymorphism, higher-order functions) and logic programming
(computing with partial information, unification, constraints).
The syntax of Curry is close to Haskell\footnote{%
Variables and function names usually
start with lowercase letters and the names of type and data constructors
start with an uppercase letter. The application of $f$
to $e$ is denoted by juxtaposition (``$f~e$'').}
\cite{PeytonJones03Haskell}. In addition, Curry allows free (logic) 
variables in conditions and right-hand sides of defining rules.
The operational semantics is based on an optimal evaluation strategy
\cite{AntoyEchahedHanus00JACM}---a conservative extension
of lazy functional programming and logic programming.

A Curry program consists of the definition of data types
(introducing \emph{constructors} for the data types) and
\emph{functions} or \emph{operations} on these types.
For instance, the data types for Boolean values and polymorphic lists
are as follows:
\begin{curry}
data Bool = False | True
data List a = [] | a : List a    -- [a] denotes "List a"
\end{curry}
A \emph{value} is an expression without defined operations.
As mentioned in Sect.~\ref{sec:Introduction},
Curry allows the definition of non-deterministic operations
with the choice operator \ccode{?} so that
the expression \ccode{True$\;$?$\;$False} has two values:
\code{True} and \code{False}.
Using non-deterministic operations as arguments
might cause a semantical ambiguity which has to be fixed.
For instance, consider the operations
\begin{curry}
xor True  x = not x            not True  = False
xor False x = x                not False = True$\listline$
xorSelf x = xor x x            aBool = True ? False
\end{curry}
and the expression \ccode{xorSelf$\;$aBool}.
If we interpret this program as a term rewriting system,
we could have the derivation
\begin{curry}
xorSelf aBool  $\to~$  xor aBool aBool     $\to~$  xor True aBool
               $\to~$  xor True False      $\to~$  not False        $\to~$ True
\end{curry}
leading to the unintended result \code{True}.
Note that this result cannot be obtained if we use a strict strategy
where arguments are evaluated prior to the function calls.
In order to avoid dependencies on the evaluation strategies
and exclude such unintended results,
Gonz\'alez-Moreno et al.\ \cite{GonzalezEtAl99} proposed
the rewriting logic CRWL as a logical
(execution- and strategy-independent) foundation for declarative
programming with non-strict and non-deterministic operations.
CRWL specifies the \emph{call-time choice} semantics \cite{Hussmann92}%
\label{ctc-semantics},
where values of the arguments of an operation are determined before the
operation is evaluated. This can be enforced in a lazy strategy
by sharing actual arguments.
For instance, the expression above can be lazily evaluated
provided that all occurrences of \code{aBool}
are shared so that all of them reduce either to \code{True} or to \code{False}
consistently.
Thus, sharing is not an option to support an efficient evaluation,
but it is required for semantical reasons.

Fortunately, sharing is not an extra burden
but already provided by implementations of lazy languages
in order to avoid duplication of work.
To avoid re-evaluations of identical subexpressions,
e.g., the subexpression \code{$f\;e$} in \code{xorSelf$\;$($f\;e$)}
where $f$ might cause an expensive computation,
the two occurrences of \code{x} in the right-hand side
of the \code{xorSelf} rule are \emph{shared}.
This can be achieved by a graph represention of expressions
so that all occurrences of \code{x} refer to the same graph node.
Hence, if \code{$f\;e$} is evaluated as the first argument of \code{xor}
to some value $v$, the node containing $f$ is replaced by $v$
so that the second argument of \code{xor} also refers to $v$.
This ``update-in-place'' of evaluated function calls is
essential for lazy languages and also required to ensure the optimality
of lazy strategies for functional logic languages
\cite{AntoyEchahedHanus00JACM}.

Formally, we can consider programs as graph rewriting systems
\cite{Plump99Handbook} so that rewrite steps are graph replacements.
In order to simplify our presentation, we use the idea to represent
sharing by \code{let} expressions, similarly
to Lauchbury's natural semantics for lazy evaluation \cite{Launchbury93}.
This is also used to specify the operational semantics of
functional logic languages \cite{AlbertHanusHuchOliverVidal05}.
Bindings of \code{let} expressions are stored in a heap
so that updates of function nodes are represented as heap updates.
Instead of repeating the details of \cite{AlbertHanusHuchOliverVidal05},
we show the possible evaluations of \code{xorSelf$\;$aBool} in this heap model.
Here, the heap is shown on the left, evaluations with the same heap are
written in the same line, and new evaluation tasks caused by non-deterministic
choices are indented:
\begin{center}
\begin{tabular}{l@{~~~~}l@{~~~~~~}r}
$[]$ & \code{let x = aBool in xorSelf x} & (1)\\
$[\code{x} \mapsto \code{aBool}]$ & $\to~ \code{xorSelf x}$ $\to~ \code{xor x x}$ & (2)\\
$[\code{x} \mapsto \code{True ? False}]$ & $\to~ \code{xor x x}$ & (3)\\
$~~~[\code{x} \mapsto \code{True}]$ & $~~~\code{xor x x} ~\to~ \code{not x} ~\to~ \fbox{\code{False}}$ & (4)\\
$~~~[\code{x} \mapsto \code{False}]$ & $~~~\code{xor x x} ~\to~ \code{x} ~\to~ \fbox{\code{False}}$ & (5)\\
\end{tabular}
\end{center}
In line (2), the \code{let} binding is moved into the heap.
The function call in this binding is evaluated and updated in (3).
Since this is a choice,
a new evaluation task is established for each alternative.
Thanks to the sharing of the value of \code{x}, the unintended
value \code{True} is not computed as a result.
Since a heap can be considered as another representation of a graph
structure, we use heaps and graphs interchangeably.

\section{Pull-Tabbing}
\label{sec:Pulltabbing}

If non-deterministic choices are implemented by backtracking, as in Prolog,
one has to reason about the influence of the search strategy
to the success of an evaluation---a non-trivial task in the presence
of a lazy evaluation strategy.
To support better search strategies,
like breadth-first or parallel search,
all non-deterministic choices should be represented in a single
(graph) structure so that one
can easily switch between different computation branches.
As discussed in Sect.~\ref{sec:Introduction},
pull-tabbing \cite{AlqaddoumiAntoyFischerReck10,Antoy11ICLP}
has been used in implementations supporting advanced search strategies.
A \emph{pull-tab step} moves a choice occurring in a
demanded argument of an operation outside this operation:
if $f$ demands the value of its single argument,
then
\[
f~(e_1~\code{?}~e_2) ~~\to~~ (f~e_1) ~\code{?}~ (f~e_2)
\]
is a pull-tab step.

The nice aspects of pull-tabbing are its operational completeness
\cite{Antoy11ICLP} and the locality of steps.
Iterated pull-tab steps move choices towards the root:
\begin{curry}
not (not (True ? False)) $\to$ not ((not True) ? (not False))
                         $\to$ (not (not True)) ? (not (not False))
\end{curry}
A choice at the root of an expression leads to two new expressions
that must be evaluated and might lead to two result values.
Since pull-tabbing does not fix some search strategy,
it is assumed that these alternative expressions are evaluated
by different computation \emph{tasks}.
Conceptually, an entire computation consists of a set of tasks
where each task evaluates some node of a graph.
In this example, the new tasks evaluate the expressions
\code{not$\;$(not$\;$True)} and \code{not$\;$(not$\;$False)}, respectively.

Pull-tab steps as described so far are not sufficient to
correctly implement a non-strict functional logic language like Curry.
As discussed in Sect.~\ref{sec:Curry},
the call-time choice semantics requires to share the values
of non-deterministic arguments in a computation.
We can implement this requirement by adding identifiers to choices and
associating a ``fingerprint'' \cite{AntoyJost16} to each task.
A \emph{fingerprint} is a (partial) mapping from choice identifiers
to choice alternatives ($L$eft or $R$ight).
When a task reaches a choice at the root, it proceeds as follows:
\begin{itemize}
\item If the fingerprint of the task contains a selection for this choice,
      select the corresponding branch of this choice.
\item Otherwise, create two new tasks for the left and right alternative
      where the fingerprint is extended for this choice with 
      $L$ and $R$, respectively.
\end{itemize}
With this refinement of pull-tabbing, we obtain the following
evaluation of a variation of \ccode{xorSelf$\;$aBool}
(where the fingerprint of the task
is written on the left and the heap, which is always
$[\code{x} \mapsto \code{True ?$_1$ False}]$, is omitted):
\begin{center}
\begin{tabular}{l@{~~~~~}l}
$[]$    & \code{xor x x $\to$ (xor True x) ?$_1$ (xor False x)} \\
$[1/L]$ & ~~~\code{xor True x $\to$ not x $\to$ (not True) ?$_1$ (not False)} \\
$[1/L]$ & ~~~$\to$ \code{not True  $\to$ \fbox{\code{False}}} \\
$[1/R]$ & ~~~\code{xor False x $\to$ x $\to$ True ?$_1$ False} \\
$[1/R]$ & ~~~$\to$ \fbox{\code{False}} \\
\end{tabular}
\end{center}
Thanks to fingerprints, only values which
are correct w.r.t.\ the call-time choice semantics are
produced \cite{Antoy11ICLP}.

Unfortunately, pull-tabbing has some performance problems.
In contrast to backtracking, where a non-deterministic choice
is implemented by selecting one branch and proceed with
this selection until failure or success,
pull-tabbing moves each non-deterministic choice up to the root
of the expression under evaluation.
Hence, the consistency of choices is checked only for choices
at the root, i.e., outside function calls.
This has the operational consequence that \emph{each} access to
a non-deterministic expression leads to a stepwise shifting
of choices towards the root.
Thus, multiple accesses to a same non-deterministic expression
multiplies the execution time.
For instance, consider a function
\begin{equation}\label{example:sharing}
f~x \code{~=~} C[x,\ldots,x]
\end{equation}
where the right-hand side is (or evaluates to) an expression
containing $n$ occurrences of the argument $x$
(represented by the context $C$).
Now consider the evaluation of $f\;(e_1~\code{?}~e_2)$.
Whenever some occurrence of $x$ in $C$ is demanded in this evaluation,
the choice occurring in the actual argument is moved up to the root by pull-tabbing.
Hence, if all $n$ occurrences of $x$ in $C$ are demanded
at depth $d$, approximately $n \cdot d$ pull-tab steps are performed
(and most of the resulting choice nodes are omitted at the end
due to fingerprinting).
In contrast, backtracking is more efficient
since it selects one alternative for the first occurrence of $x$
and then simply uses this alternative for all subsequent
occurrences of $x$.

\section{Memoized Pull-Tabbing}
\label{sec:MPT}

In this section we present an improvement of pull-tabbing
which avoids the performance problems discussed above.

\subsection{The Basic Scheme}

In principle, the duplication of choices is necessary
due to shared subexpressions
which are evaluated by different tasks.
In contrast to purely functional programming,
it would be wrong to update graph nodes of such expressions
by their results if they occur in a non-deterministic context
(this is discussed in more detail in Appendix~\ref{app:wrong-pull-tabbing}).
As a solution to this problem,
we propose to store ``task-specific'' updates in the graph,
i.e., instead of updating graph nodes by their computed results,
we keep the graph nodes but memorize results already computed by some task
for a function node.
When a task has to evaluate a function node again due to sharing,
it directly uses an already computed result.

In order to implement this idea,
each task evaluating some expression (subgraph)
has a unique identifier (e.g., a number), also called
\emph{task identifier}.
To store task-specific updates,
each graph node representing a function call
contains a (partial) map $tr$, called
\emph{task result map}, from task identifiers to graph nodes.

To avoid repeated pull-tab steps,
pull-tab steps are not performed for choices that
already contain a selection in the fingerprint of the task.
In this case, we proceed with the selected branch
but have to remember, by using the task result map,
that computed results are valid only for this branch.
To be more precise, consider a node $n$
representing some function call $f\;(e_1\;\code{?}_{c}\;e_2)$,
where $f$ demands its argument.
This node is evaluated by the task with identifier $i$ as follows:
\begin{itemize}
\item
If task $i$ contains no selection for $c$ in its fingerprint,
a standard pull-tab step is performed, i.e.,
two new nodes $n_1 = f\;e_1$ and $n_2 = f\;e_2$ are created
and $n$ is updated to $n_1 \;\code{?}_{c}\; n_2$.
\item
If task $i$ contains a selection for $c$, say $L$,
it would be wrong to update node $n$ to $f\;e_1$
due to possible sharing of $n$
(see Appendix~\ref{app:wrong-pull-tabbing} for an example).
Instead, a new node $n' = f\;e_1$ is created and
$n.tr$ is updated with $n.tr(i) = n'$.
\end{itemize}
This strategy has the consequence that only the first occurrence of a choice
in a computation is moved to the root by iterated pull-tab steps.
Since a choice at the root causes a splitting of the current task
into two new tasks evaluating the left and right alternative,
respectively, this choice, when evaluated again due to sharing,
has a selection in the fingerprint so that this selection is
immediately taken and stored in the task result map.

Since function calls can be nested, the task result map
must be considered for any function call, i.e., also those without
a choice in an argument.
Thus, when a task with identifier $i$ evaluates
some function node $n$, it checks whether $n.tr(i)$ is defined:
\begin{itemize}
\item
If $n.tr(i) = n'$, then $n'$ is evaluated instead of $n$.
\item
If $n.tr(i)$ is undefined, $n$ is evaluated to some node $n'$ and
$n.tr$ is updated with $n.tr(i) = n'$.
\end{itemize}
Hence, if node $n$ is shared so that task $i$ has to evaluate $n$ again,
the already evaluated result is taken.

This evaluation scheme requires a bit more time when function nodes are accessed
but avoids the expensive duplication of non-deterministic computations
with pure pull-tabbing.
We call this improved strategy \emph{memoized pull-tabbing} (\emph{MPT}).

Memoized pull-tabbing can reduce the complexity of non-deterministic
computations. For instance, consider again function $f$
defined by rule~(\ref{example:sharing}) in Sect.~\ref{sec:Pulltabbing}.
When the expression \code{let x = $e_1\;$?$_1\;e_2$ in $f\;$x} is evaluated,
rule~(\ref{example:sharing}) is applied and eventually
the first occurrence of \code{x} is evaluated by a pull-tab step.
This leads (by iterated pull-tab steps) to the expression
\begin{curry}
let x = $e_1\;$?$_1\;e_2$ in $C[e_1,$x$,\ldots,$x$]$ ?$_1$ $C[e_2,$x$,\ldots,$x$]$
\end{curry}
The left and right alternative are further evaluated by
two tasks $T_1$ and $T_2$
having an $L$ and $R$ selection for choice $1$, respectively.
Task $T_1$ evaluates all further occurrences of \code{x}
by selecting $e_1$ and setting the task result maps of the
parent nodes to the results computed with this selection.
Hence, instead of $n \cdot d$ pull-tab steps with pure pull-tabbing,
MPT performs only $d$ pull-tab steps and $n-1$ ``selection'' steps
for each task $T_1$ and $T_2$ to evaluate the initial expression.

Before presenting and evaluating an implementation of MPT,
we propose some refinements which will lead to our final
MPT strategy.

\subsection{Refinements for Deterministic Computations}

In typical application programs, large parts of an evaluation are
\emph{deterministic computations}, i.e., computations
where choice nodes do not occur.
Similarly, a \emph{deterministic expression} is an expression
whose evaluation does not demand the evaluation of a choice.
Since a reasonable implementation of a functional logic language
should support efficient deterministic computations,
we present two improvements of our basic MPT strategy for this purpose.

Our first refinement tries to avoid the use of the
task result map $tr$ in nodes whenever it is not necessary,
in particular, for deterministic computations.
For this purpose, each graph node $n$ has an
\emph{owner task} (\emph{ot}),
i.e., $n.ot$ is the identifier of the task that created this node.
For the initial expression, the owner task of all nodes
is the identifier of the main task.
When a rule is applied, i.e., a function call is replaced
by the right-hand side of a program rule, the owner task
of the nodes created for the right-hand side is identical
to the owner task of the root of the left-hand side.
In case of a pull-tab step
\[
f~(e_1~\code{?}~e_2) ~~\to~~ (f~e_1) ~\code{?}~ (f~e_2)
\]
the owner tasks of the new
function calls $f\;e_1$ and $f\;e_2$
are the identifiers of the new tasks that will evaluate
the left and right alternative, respectively.

In order to compare the owner tasks of nodes, we assume a partial ordering
on task identifiers.
Note that new tasks are created when a choice appears at the root.
In this case the current task $t$ is split into two new tasks $t_1,t_2$
which evaluate the left and right alternative of the choice, respectively.
We call $t$ \emph{parent} of $t_1$ and $t_2$.
If $i,i_1,i_2$ are the identifiers of $t,t_1,t_2$, respectively,
then we assume that $i < i_1$ and $i < i_2$.
We call node $n_1$ \emph{younger} than $n_2$ if $n_1.ot > n_2.ot$.

If the current task evaluates some choice $n = e_1\;\code{?}_c\; e_2$
and the fingerprint of the task already contains a decision
for this choice, we follow this decision instead of pushing
the choice towards the root by a pull-tab step, as described above.
Now we refine the basic scheme by considering the owner tasks.
Assume that $i$ is the identifier of the current task,
its fingerprint selects $e_1$ for choice $c$,
and there is the parent node $n' = f\;n$
(for simplicity, we consider only unary functions in this discussion).
We distinguish the following cases:
\begin{enumerate}
\item
If $i > n'.ot$, then $n'.tr(i)$ is set to a new node $n'' = f\;e_1$
with $n''.ot = i$
and the evaluation proceeds with node $n''$.
\item
Otherwise ($i = n'.ot$), node $n'$ is updated in place
such that $n' = f\;e_1$.
\end{enumerate}
Next consider the situation that there is some function node $n = f\;a$
and the argument $a$ has been evaluated to $a'$.
\begin{enumerate}
\item 
If $a'$ is younger than $n$,
the argument has been evaluated to some value which is valid only
in the new task which created $a'$.
Instead of updating $n$ in place,
$n.tr(a'.ot)$ is set to a new node $n' = f\;a'$
with $n'.ot = a'.ot$ and the evaluation proceeds with
node $n'$ instead of $n$.
\item
Otherwise ($a'$ is not younger than $n$),
the value computed for $a$ is valid for $n$ so that $n$
is updated in place such that $n = f\;a'$.
\end{enumerate}
Thus, for deterministic computations, which are performed in a single
task (tasks are created only for non-deterministic steps),
the task result maps are not used at all.

A further refinement exploits the tree structure of tasks.
An \emph{ancestor} of a task is either its parent or the parent
of some ancestor of the task.
Consider the case that we have to evaluate some function node $n$
in a task identified by $i$:
\begin{enumerate}
\item 
If $n.tr(i)$ is defined, then task $i$ already evaluated node $n$
so that we can proceed with $n.tr(i)$ instead of $n$.
\item
If $n.tr(i)$ is not defined and $j$ is a parent of $i$ so that
$n.tr(j) = n'$, then $n'$ is also a valid result of $n$
so that we can proceed with $n'$.
Hence, we follow the ancestor chain of $i$ to find the first ancestor $k$
such that $n.tr(k)$ is defined.
\item
Otherwise (there is no ancestor $j$ of $i$ with $n.tr(j)$ defined),
we evaluate $n$.
\end{enumerate}
Note that the owner task of nodes changes in a computation
only if some choice node is evaluated.
In case of a pull-tab step, the new nodes are younger than the
choice node.
If there is some decision for the choice w.r.t.\ the fingerprint
of the current task, we commit to the selected branch and
set the owner task of this selection to the current task.
An interesting consequence of this strategy
is that the owner tasks of nodes in a computation
are not changed when choice nodes do not occur in this computation.
In particular, deterministic computations without occurrences
of choice nodes are always evaluated in place---independently of
the task which evaluated them.
This has the effect that deterministic expressions
are evaluated at most once, even if they are shared among
non-deterministic branches.
This property, also called
\emph{sharing across non-determinism} \cite{BrasselHuch07},
is an important feature of the pull-tab strategy.
Consider an expression
\begin{curry}
let $x$ = $e$ in $C_1[x]$ ? $C_2[x]$
\end{curry}
where $e$ is a deterministic expression
and the value of $x$ is demanded in both $C_1[x]$ and $C_2[x]$.
Then $e$ will be evaluated only once since the task
evaluating $C_1[x]$ will replace $e$ by its result
so that this result is available for the task evaluating $C_2[x]$.
In contrast, an implementation based on backtracking
would evaluate $e$ two times since the evaluation of $e$ by
the task evaluating $C_1[x]$ will be undone before evaluating $C_2[x]$.
Note that this is not only a problem of backtracking.
For instance, the approach to implement call-time choice
with purely functional programming features presented
in \cite{FischerKiselyovShan11} also reports the lack of
sharing across non-determinism.

\section{Implementation}
\label{sec:implementation}

In order to evaluate our ideas, we implemented MPT
in Julia\footnote{\url{https://julialang.org/}},
a high-level dynamic programming language.
We used Julia due to its direct support of dynamic data structures,
garbage collection, and higher-order features.
By exploiting the intermediate language ICurry \cite{AntoyHanusJostLibby20},
the compiler from ICurry to Julia is approximately 300 lines of Curry code.
Furthermore, the run-time system, responsible to implement
the computation graph, pull-tab steps, computation tasks
with various search strategies, and some more aspects,
consists of approximately 300 lines of Julia code.
Thus, this implementation, called ``JuCS'' (Julia Curry System\footnote{%
Available at \url{https://github.com/cau-placc/julia-curry}}),
is a proof of concept which could also be implemented,
with more effort and probably more efficiently,
in other imperative languages, like C.
In the following, we sketch some aspects of this implementation.

Apart from the memoized pull-tabbing strategy,
the implementation has many similarities to Sprite \cite{AntoyJost16}.
Expressions are represented as a graph structure.
To distinguish different kinds of graph nodes
(function, constructor, choice, failure, etc),
each node has a tag.
Furthermore, a node contains an integer value (choice identifier,
constructor index, etc), the identifier of the owner task,
an array of references to argument nodes,
a code reference in case of function nodes,
and a task result map (a Julia dictionary with task identifiers as keys
and node references as values).
The run-time system works on a queue of tasks where
each task contains a unique number,
the root node evaluated by this task,
the fingerprint, and the identifiers of the parent tasks.
With these data structures, the run-time system
evaluates expressions, as described above,
by computing the head normal form of the root node of the current task.
If this yields a choice node, two new tasks are created
with extended fingerprints.
In case of depth-first search, these tasks are added
at the front of the task queue, while they are added at the back
in case of breadth-first search.

Free variables and their bindings require
a non-trivial implementation with non-deterministic value
generator operations in a pure pull-tabbing implementation
\cite{BrasselHanusPeemoellerReck13PADL}.
Our MPT strategy allows a much simpler implementation.
Instead of representing free variables as value generators
(as in \cite{AntoyHanus06ICLP}),
JuCS has a ``free'' tag for nodes where the
task result map is used to store task-specific bindings
for free variables.
Hence, free variables are handled as efficient as in Prolog implementations
while still allowing more flexible search strategies.

In order to compare the different run-time models
(MPT, pull-tabbing, backtracking)
inside our implementation, JuCS contains also two alternative
run-time systems implementing pure pull-tabbing and backtracking.
The pull-tabbing system is a reduced variant of the standard
run-time system.
The backtracking system uses ideas from Prolog implementations,
in particular, the improved backtracking and trailing mechanism of
Warren's Abstract Machine \cite{Warren83} to reduce the amount of
stored backtrack information.

\section{Benchmarks}
\label{sec:benchmarks}

Memoized pull-tabbing requires more effort at run time
than pure pull-tabbing due to the tests when evaluating
or updating a function node in the computation graph.
Thus, it is interesting to see whether this pays off in practice.
Therefore, we executed a set of benchmarks
with our new implementation and compared the execution times\footnote{%
All benchmarks were executed on a Linux machine running
Ubuntu 18.04 with an Intel Core i7-85550U (1.8GHz) processor.}
of the different run-time systems provided with this implementation.
The results are summarized in Table~\ref{table:benchjuliarts}.

\begin{table}[t]
\begin{center}
\begin{tabular}{|r|r|r|r|}
\hline
Program & ~~~~MPT & ~~pull-tab & backtrack \\
\hline
\code{nrev} & 2.37 s & 2.29 s & 7.09 s \\
\code{takPeano} & 12.04 s & 11.84 s & 31.78 s \\
\code{addNum2} & 0.46 s & 6.21 s & 0.39 s \\
\code{addNum5} & 1.61 s & 47.69 s & 1.26 s \\
\code{select50} & 0.05 s & 4.18 s & 0.14 s \\
\code{select75} & 0.10 s & 24.10 s & 0.31 s \\
\code{select100} & 0.18 s & 111.47 s & 0.55 s \\
\hline
\end{tabular}
\end{center}
\caption{Evaluating different run-time systems of JuCS\label{table:benchjuliarts}}
\end{table}

The first two examples\footnote{%
The actual programs are available with the implementation
described in Sect.~\ref{sec:implementation}.}
are purely deterministic programs:
\code{nrev} is the quadratic naive reverse algorithm
on a list with 4096 elements and
\code{takPeano} is a highly recursive function on naturals \cite{Partain93}
applied to arguments (24,16,8), where numbers and arithmetic operations
are in Peano representation.
\code{addNum2} and \code{addNum5} non-deterministically choose
a number (out of 2000) and add it two and five times, respectively.
\code{select$n$} non-deterministically selects an element
in a list of length $n$ and sums up the element and the list
without the the selected element.

As one can see from the direct comparison to
pure pull-tabbing, the overhead caused by the
additional checks required for memoization is limited and
immediately pays off when non-deterministic expressions are shared,
which is often the case in applications involving
non-determinism.

It is interesting to note that the backtracking strategy
is less efficient than MPT, although MPT supports
more flexible search strategies. This might be due to the fact
that backtracking has to check, for each reduction step,
whether this step has to be remembered in order to undo it
in case of backtracking.

As already discussed, backtracking has another disadvantage:
deterministic computations are not shared across non-deterministic
computations. This has the unfortunate consequence that
a client using some algorithm of a library has to know
whether this algorithm is implemented with non-deterministic
features, since arguments might be evaluated multiple times
with backtracking.
To show this effect,
consider the deterministic insertion sort operation \code{isort},
the non-deterministic permutation sort operation \code{psort},
and the infinite list of all prime numbers \code{primes}
together with the following definitions (this example
is inspired from \cite{BrasselHanusPeemoellerReck11}):
\begin{curry}
sort1 = isort$\;$[primes!!303, primes!!302, primes!!301, primes!!300]
sort2 = psort$\;$[primes!!303, primes!!302, primes!!301, primes!!300]
\end{curry}
In principle, one would expect that the execution time
of \code{sort1} is almost equal to the time to execute
\code{sort2} since the time to sort a four-element list is neglectable.
However, implementations based on backtracking
evaluate the primes occurring in \code{sort2} multiple times,
as can be seen by the run times shown
in Table~\ref{table:bench-sharing-across-nondet}.
\begin{table}[t]
\begin{center}
\begin{tabular}{|r|r|r|r|}
\hline
Program & ~~~MPT & ~~pull-tab & backtrack \\
\hline
\code{sort1} & 9.83 s & 9.96 s &  15.47 s \\
\code{sort2} & 9.84 s & 9.73 s & 155.64 s \\
\hline
\end{tabular}
\end{center}
\caption{Effect of sharing across non-determinism\label{table:bench-sharing-across-nondet}}
\end{table}

We already emphasized the fact that pull-tabbing supports
flexible search strategies. Since all non-deterministic values
of an expression are represented in one structure,
different search strategies can be implemented
as traversals on this structure.
For instance, KiCS2 evaluates each expression to a tree
of its values so that the top-level computation collecting all values
can be defined as a traversal on this tree \cite{BrasselHanusPeemoellerReck11}.
Our implementation uses a queue of tasks so that
search strategies can be implemented as
specific strategies to put and get tasks to and from the queue, respectively
(as sketched above).
The practical behavior of search strategies in KiCS2
was analyzed in \cite{HanusPeemoellerReck12ATPS} (since then,
breadth-first search became the default strategy for KiCS2).
Table~\ref{table:bench-dfs-vs-bfs} shows that MPT has an even
better behavior since, in contrast to pure pull-tabbing,
it is not necessary to move all choices to the root in order
to build a tree of all values.
Here, we also added the classical permutation sort example
since it showed a larger slowdown of breadth-first search
in \cite{HanusPeemoellerReck12ATPS}.

\begin{table}[t]
\begin{center}
\begin{tabular}{|r|r|r|}
\hline
Example & KiCS2 (BFS/DFS) & JuCS (BFS/DFS) \\
\hline
\code{addNum2} & 2.18 & 1.02 \\
\code{addNum5} & 1.58 & 1.00 \\
\code{select50} & 1.03 & 1.00 \\
\code{select100} & 1.28 & 1.04 \\
\code{permsort} & 4.46 & 1.11 \\
\hline
\end{tabular}
\end{center}
\caption{Relative execution times of BFS vs.\ DFS\label{table:bench-dfs-vs-bfs}}
\end{table}

\section{Related Work}
\label{sec:related}

In this section we review other approaches to implement
functional logic languages and relate our proposal to them.

Early approaches to implement functional logic languages
exploited Prolog's backtracking strategy for non-deterministic
computations
\cite{AntoyHanus00FROCOS,LoogenLopezFraguasRodriguezArtalejo93PLILP}.
By adding a mechanism to implement demand-driven evaluation,
one can use Prolog as a target language, as done in
PAKCS \cite{Hanus18PAKCS} and TOY \cite{Lopez-FraguasSanchez-Hernandez99}.
The usage of Prolog yields also a direct support for free variables.
However, such implementations suffer from the
operational incompleteness of the backtracking strategy.

Pull-tabbing supports more flexible search strategies by
representing choices as data.
The theoretical properties are investigated in \cite{Antoy11ICLP}.
On the practical side, pull-tabbing is useful to implement
non-determinism in a deterministic target language. 
For instance, ViaLOIS \cite{AntoyPeters12FLOPS} uses pull-tabbing to translate
Curry programs into Haskell and OCaml programs, respectively.
ICurry \cite{AntoyHanusJostLibby20} is an intermediate language
intended to translate Curry programs into imperative target languages.
It has been used to translate Curry to LLVM code \cite{AntoyJost16}
and to C or Python programs \cite{Wittorf18}.
The operational semantics of ICurry is specified
in \cite{AntoyHanusJostLibby20} by an abstract machine
which performs pull-tab steps and
uses a graph structure to represent expressions with sharing
and a queue of tasks, where each task has its own fingerprint
to implement the selection of consistent choices.

The Curry compiler KiCS2 \cite{BrasselHanusPeemoellerReck11}
is based on pull-tabbing and compiles Curry programs into a purely
functional Haskell programs.
Non-determinism is implemented by representing
choices as data terms so that
Curry expressions are evaluated to choice trees.
Pull-tab steps are encoded as rules for choice terms.
Values are extracted from choice trees
by traversing them with fingerprints.
Hence, KiCS2 implements non-determinism in a modular way:
any expression is evaluated to a choice tree representation of all its values,
and there is a separate operation which extracts correct values
from the choice tree structure.
Therefore, KiCS2 implements various search strategies
as different tree traversal strategies,
and infinite search spaces (choice trees) do not cause problems
due to Haskell's lazy evaluation strategy.

Since KiCS2 suffers from the performance problems of pure pull-tabbing
(see Appendix~\ref{app:comparison-of-curry-systems} for some benchmarks),
an eager evaluation of demanded non-deterministic subexpressions
is proposed in \cite{Hanus12ICLP}.
An automatic program transformation implementing this optimization
is based on a demand analysis.
However, this approach does not work for arbitrary programs
since a precise demand analysis for complex data structures
is non-trivial and not yet available for functional logic programs.
Therefore, it is an interesting question for future work
whether our MPT scheme can be combined with the
purely functional implementation approach of KiCS2.

\emph{Sharing across non-determinism} describes the property
that deterministic subexpressions shared in different
non-deterministic branches are evaluated at most once.
This is usually not the case in implementations based on backtracking.
As emphasized in \cite{BrasselHuch07},
pull-tabbing easily ensures this property if the target
language implements sharing for common subexpressions,
as in our implementation or in lazy functional target languages
\cite{BrasselHanusPeemoellerReck11,BrasselHuch07}.

An approach to support functional logic programming features
as in Curry in a purely functional language is a
library for non-deterministic computations with (explicit) sharing
\cite{FischerKiselyovShan11}.  The key idea of this library is to
translate non-deterministic computations into monadic computations
that manipulate a \emph{thunk store}.  The thunk store holds either
unevaluated computations or their results, which may again contain
unevaluated arguments, and is closely related to the heap described in
Sect.~\ref{sec:Curry}.  The library provides an explicit \code{share}
operation to allow the sharing of computations.  Shared computations
are initially entered unevaluated into the thunk store and only the
demand for a computation triggers its evaluation.  If a computation is
non-deterministic, the thunk store is updated with the corresponding
result independently in each branch.  All subsequent uses of the
shared computation within one computation branch then reuse the
updated result in the thunk store.  Although shared results are reused
in one computation branch if demanded more than once, the library does
not support sharing across non-determinism because shared computations
are evaluated independently in different branches.  Due to the fact
that the implementation relies on the type class \code{MonadPlus},
different search strategies can be exploited depending on the concrete
instance of \code{MonadPlus}.  Furthermore, the library has no direct
support of free variables and can only emulate them by using
non-deterministic generators \cite{AntoyHanus06ICLP}.

Table~\ref{table:comparison} compares the properties of
the various approaches to implement demand-driven non-deterministic
computations discussed above.
``Flexible search strategies'' means whether only one
or a number of different search strategies are supported.
``Free variables'' denotes a direct support of free variables.
This is not the case for explicit sharing and pull-tabbing,
since they require a simulation of free variables by
non-deterministic generator operations and non-trivial techniques
to obtain the effect of binding free variables through unification
\cite{BrasselHanusPeemoellerReck13PADL}.
``Sharing across non-determinism'' describes the aforementioned ability
to reuse already computed results of deterministic subexpressions
in different branches of non-deterministic computations.
``Sharing non-determinism'' means that the results of already evaluated
non-deterministic subexpressions are re-used when these subexpressions
are shared.
As one can see, our new MPT strategy is the only implementation
which combines all these properties.
As shown by our benchmarks, this has a positive effect
on the efficiency of MPT on a range of different application scenarios.

\begin{table}[t]
\begin{center}
\begin{tabular}{lcccc}
 & \parbox{1.7cm}{\centering Back-\\ tracking}
 & \parbox{1.7cm}{\centering Explicit\\ Sharing}
 & \parbox{1.7cm}{\centering Pull\\ Tabbing} & MPT \\
\hline
Flexible search strategies      & $-$ & $+$ & $+$ & $+$ \\
Free variables                  & $+$ & $-$ & $-$ & $+$ \\
Sharing across non-determinism  & $-$ & $-$ & $+$ & $+$ \\
Sharing non-determinism         & $+$ & $+$ & $-$ & $+$ \\
\end{tabular}
\end{center}
\caption{Comparing properties of implementation strategies%
\label{table:comparison}}
\end{table}

\section{Conclusions}
\label{sec:Conclusion}

The efficient implementation of functional logic programming languages
is still a challenge due to the combination of advanced
declarative programming concepts.
In order to free the programmer from considering details
about the concrete evaluation strategy, it is desirable to support
operationally complete strategies which ensure that
values are computed whenever they exist.
This can be obtained by representing the complete state
with all branches of a non-deterministic computation
in one data structure.
Pull-tabbing is a simple and local transformation to deal with
non-deterministic choices.
However, pull-tabbing has the risk to duplicate work during evaluation.
In this paper we proposed a significant improvement
by adding a kind of memoization to pull-tabbing.
As we demonstrated by our benchmarks,
this improved evaluation mechanism does not cause much overhead,
is often faster than backtracking, and can dramatically improve
pure pull-tabbing. Morever, it keeps all the positive
properties of pull-tabbing:
application of various search strategies and
sharing across non-determinism.
Our prototypical implementation showed that it can also be implemented
with modest efforts: Curry programs can be compiled by using
the already existing intermediate language ICurry in a straightforward
manner, and the run-time system is quite compact.
Thus, it is an ideal model to implement
multi-paradigm declarative languages also with other target
languages, e.g., to integrate declarative programming
in applications written in other imperative languages.

Nevertheless, there is room for future work.
For instance, one could try to identify non-shared subexpressions,
e.g., by some sharing or linearity analysis, to avoid
run-time checking of memoized data.
Another interesting question is whether it is possible to
implement the presented ideas in a purely functional manner
so that one can use them to improve existing approaches
like KiCS2 \cite{BrasselHanusPeemoellerReck13PADL} or
the library for explicit sharing \cite{FischerKiselyovShan11}.

\bibliographystyle{plain}
\bibliography{paper}

\newpage
\appendix

\section{Improve Pull-Tabbing: The Wrong Way}
\label{app:wrong-pull-tabbing}

This appendix discusses why pull-tabbing duplicates choice nodes
even if the task has already made a selection for this choice
(stored in the fingerprint).
To be more precise, we will discuss the following question:
\begin{quote}
Is it possible to use an already selected choice alternative at any place
and not only at the root?
\end{quote}
For a rule like
\begin{curry}
$f~x$ = $C[x,\ldots,x]$
\end{curry}
this means that when the argument $x$ is bound to a non-deterministic
expression and some branch is selected for the first occurrence of $x$
in the right-hand side, the same selection for all
further occurrences is used instead of pull-tabbing their choices to the root.

Unfortunately, this is unsound due to sharing of
non-deterministic expressions.
As an example, consider the expression
\begin{curry}
let { x = False$\;$?$\;$True ; y = not x } in (x && y) ? y
\end{curry}
With call-time choice,
the left alternative \code{(x$\;$\&\&$\;$y)}
evaluates to the values \code{False} and \code{False}
and the right alternative \code{y} evaluates to \code{True} and \code{False}.
These values are also obtained with pull-tabbing.

Now consider what happens if we try to improve pull-tabbing
by using already selected choices whenever these choices occur
in a computation.
Figure~\ref{fig:pull-tab-wrong} shows the derivation
of all values with this modified pull-tabbing strategy.

\begin{figure}[ht]
\begin{center}
\begin{tabular}{l@{~~}l@{~~~}lr}
Fingerprint: & Heap: & \hspace{-2ex}Expression: \\ \hline
$[]$          & $[\code{x} \mapsto \code{True$\;$?$_1\;$False}, \code{y} \mapsto \code{not x}]$ & \hspace{-2ex}\code{(x \&\& y) ?$_2$ y} & (1)\\
$[2/L]$       & $[\code{x} \mapsto \code{True$\;$?$_1\;$False}, \code{y} \mapsto \code{not x}]$ & \code{(x \&\& y)} & (2)\\
$[2/L]$       & $[\code{x} \mapsto \code{True$\;$?$_1\;$False}, \code{y} \mapsto \code{not x}]$ &
              \code{(True \&\& y)$\;$?$_1\;$(False \&\& y)} & (3) \\
$[2/L, 1/L]$  & $[\code{x} \mapsto \code{True$\;$?$_1\;$False}, \code{y} \mapsto \code{not x}]$ & ~~\code{(True \&\& y) $\to$ y} & (4)\\
$[2/L, 1/L]$  & $[\code{x} \mapsto \code{True$\;$?$_1\;$False}, \code{y} \mapsto \code{not True}]$ & ~~\code{y} & (5)\\
$[2/L, 1/L]$  & $[\code{x} \mapsto \code{True$\;$?$_1\;$False}, \code{y} \mapsto \code{False}]$ & ~~\code{y $\to$ \fbox{False}} & (6)\\
$[2/L, 1/R]$  & $[\code{x} \mapsto \code{True$\;$?$_1\;$False}, \code{y} \mapsto \code{False}]$ & ~~\code{(False \&\& y) $\to$ \fbox{False}} & (7)\\
$[2/R]$       & $[\code{x} \mapsto \code{True$\;$?$_1\;$False}, \code{y} \mapsto \code{False}]$ & \code{y $\to$ \fbox{False}} & (8) \\
\end{tabular}
\end{center}
\caption{Pull-tabbing with unrestricted selection of alternatives%
\label{fig:pull-tab-wrong}}
\end{figure}

In line (5), the choice for the argument \code{x} of \code{not}
is reduced to \code{True} due to the fingerprint of the task.
As a consequence, \code{y} is updated to \code{False}
so that the alternative value \code{True} for \code{y}
is lost in the subsequent evaluation of the task with fingerprint
$[2/R]$.
Altogether, we obtain the values \code{False}, \code{False}, and \code{False}
with this ``improvement.''

\newpage
\section{Comparison with other Curry Systems}
\label{app:comparison-of-curry-systems}

In order to get some idea of the efficiency of JuCS with
other existing major Curry implementations,
we compared JuCS with its MPT strategy and the Curry implementations
PAKCS (based on backtracking) and KiCS2 (based on pull-tabbing).
Table~\ref{table:benchpakcskics} shows the run times
(in seconds as the average of three runs)
of the examples of Table~\ref{table:benchjuliarts}
executed with JuCS, PAKCS, and KiCS2.
PAKCS compiles Curry into Prolog where
the table shows the results of two different Prolog back ends:
SICStus-Prolog (Version 4.3) and SWI-Prolog (Version 7.6).
KiCS2 compiles Curry into Haskell (GHC Version 8.4)
where non-determinism is implemented by pull-tabbing.

This table shows that KiCS2 is quite fast on deterministic
computations (due to the efficient Haskell implementation provided by GHC),
but it might dramatically slow down on non-deterministic
computations due to pull-tabbing.
Although JuCS is not as efficient as KiCS2
on deterministic programs (but comparable to PAKCS with its
fastest back end), it is more efficient than KiCS2
on non-deterministic examples
due to the memoized pull-tabbing strategy.

\begin{table}[t]
\begin{center}
\begin{tabular}{|r|r|r|r|r|}
\hline
Example & JuCS (MPT) & PAKCS (SICStus) & PAKCS (SWI) & KiCS2 (GHC) \\
\hline
\code{nrev} & 2.37 & 3.70 & 20.55 & 0.31 \\
\code{takPeano} & 12.04 & 16.05 & 130.87 & 0.32 \\
\code{addNum2} & 0.46 & 1.98 & 0.39 & 1.66 \\
\code{addNum5} & 1.61 & 2.00 & 0.45 & 7.67 \\
\code{select50} & 0.05 & 1.81 & 0.27 & 0.55 \\
\code{select100} & 0.18 & 1.87 & 1.02 & 10.40 \\
\code{select150} & 0.46 & 1.96 & 2.56 & 47.89 \\
\hline
\end{tabular}
\end{center}
\caption{Comparing JuCS with PAKCS and KiCS2\label{table:benchpakcskics}}
\end{table}

\end{document}